%% file: paper.tex
\documentclass[5p,twocolumn,times,nonatbib]{elsarticle}
\usepackage{graphicx}
\usepackage{amsmath}   
\usepackage{etoolbox}
\usepackage{amssymb}
\usepackage{epsfig}
\usepackage{lineno, hyperref}
\usepackage{url}
\usepackage[export]{adjustbox}
\usepackage[font=footnotesize,labelfont=bf,skip=5pt]{caption} 
\usepackage{subcaption}
\usepackage{xpatch}
\usepackage{xspace}
\usepackage{float}
\usepackage{tikz}
\makeatletter
\let\c@author\relax
\makeatother
\usepackage[backend=biber,style=phys,eprint=true,hyperref=true,biblabel=brackets,block=ragged]
{biblatex}

\usepackage{hyperref}
\hypersetup{
    colorlinks=true,
    linkcolor=blue,
    filecolor=magenta,      
    urlcolor=cyan,
    pdftitle={proANUBIS Construction},
    pdfpagemode=FullScreen,
    }
\def\pmbanner{}
\input{QC_paper/style}

\begin{document}
\begin{frontmatter}
\title{ \pmbanner {Construction of \proanubis:\\ A proof-of-concept detector for the ANUBIS experiment}\\%[0.8em]
\includegraphics[width=0.1\linewidth]{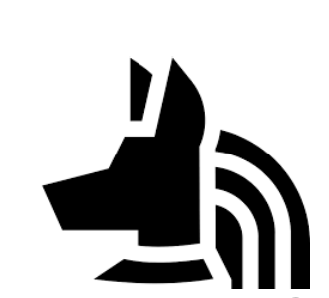}
\\
%\vspace{-7mm}
{\large\textbf{ANUBIS Collaboration}}
\\[-0.2em]
{\normalsize\textit{E-mail: }\href{mailto:anubis-publications@cern.ch}{anubis-publications@cern.ch}}
\\[-0.2em]
}
\include{QC_paper/authorList}
\begin{abstract}
The ANUBIS experiment aims to search for long-lived particles at the Large Hadron Collider (LHC) at CERN. 
To assess the feasibility of the project, a prototype detector, \proanubis, was designed, constructed, and prepared for installation in the UX1 ATLAS experimental cavern at the LHC. 
The primary physics goals of \proanubis are to determine the technical limitations of the detector technology and to explore the ANUBIS detector concept through in-situ measurements of muon and hadron fluxes inside the ATLAS cavern, which can be used to refine Monte Carlo simulations of such fluxes further.
This report describes the design and construction of the \proanubis experimental setup using Resistive Plate Chambers (RPCs), highlighting the possible future use case of the technology for ANUBIS. 
Details on the RPC technology, construction processes, quality control measures, and performance studies are discussed. 
Furthermore, the RPC front-end on-detector electronics and data acquisition components of \proanubis are presented.
\begin{center}
    %Version 2.2\\[-0.3em]
    \vspace{5mm}
    \today 
\end{center}
\end{abstract}

\end{frontmatter}

%%%%%%%%%%%%%%%
\section{Introduction} \label{Sec:Intro}

\begin{figure}[ht]
\centering
\begin{tikzpicture}
    \node[inner sep=0] (image) at (0,0) {\includegraphics[width=6.3 cm]{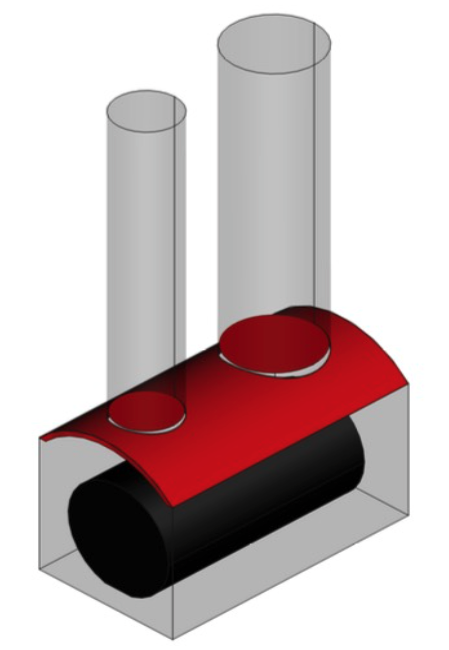}};
    \node[inner sep=0] at (0.7,3.9) {{\color{red}{\textbf{PX14}}}};
    \node[inner sep=0] at (-1.1,2.95) {{\color{red}{\textbf{PX16}}}};
    \node[inner sep=0, rotate=25] at (0.7,-3.2) {{\color{red}{\textbf{UX1}}}};
    \node[inner sep=0, rotate=-25] at (-2.1,-4.) {{\color{red}{\textbf{Side C}}}};
    \node[inner sep=0, rotate=-25] at (3.2,-1.5) {{\color{red}{\textbf{Side A}}}};
 \end{tikzpicture}

\caption{The layout of the underground cavern at Point 1 of the LHC, featuring the ATLAS experiment represented in black. 
Additionally, the PX14 and PX16 access shafts are shown. The red-coloured area illustrates the current configuration of the ANUBIS experiment, to be positioned on the ceiling of the ATLAS UX1 underground cavern.
} 
\label{fig:SketchATLAS_undergroundCavern}
\end{figure}
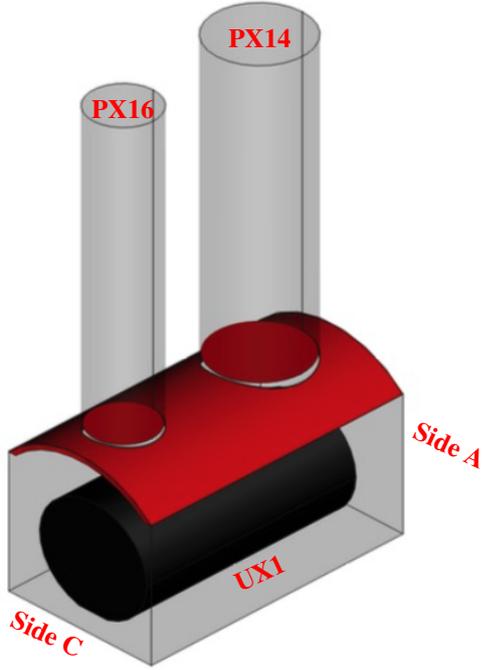

The ANUBIS\footnote{Historically known as AN Underground Belayed In-Shaft Search experiment.} experiment~\cite{Bauer:2019vqk,bib:anubis_v2} was conceived to complement and expand upon search programs for long-lived particles (LLPs)~\cite{Alimena:2019zri} at the Large Hadron Collider (LHC) at CERN~\cite{Evans:1129806}. 
As a transverse experiment, it targets models with massive LLPs with decay lengths of $\gtrsim\mathcal{O}(1~\metre)$ that are produced at partonic centre-of-mass energies at the electroweak scale and above.
The core concept of the ANUBIS experiment is to exploit the air-filled decay volume between the ATLAS detector and the ceiling of the UX1 ATLAS cavern~\cite{Bauer:2019vqk}. 
Besides dramatically reducing backgrounds, this concept minimises civil engineering costs by leveraging the existing infrastructure of the ATLAS experiment at the LHC~\cite{Aad:1129811}. 
The project's optimised configuration entails the installation of tracking detectors onto the ceiling of the ATLAS cavern and at the bottom of the PX14 and PX16 service shafts, as highlighted in red in Figure~\ref{fig:SketchATLAS_undergroundCavern}. 
Notably, the tracking detectors at the bottom of the shafts would be temporarily removed during extended technical LHC shutdowns of a couple of months, such as the Year End Technical Stops (YETS), to facilitate access to the ATLAS cavern.

This document focuses primarily on the design and construction of \proanubis, a prototype detector to showcase the feasibility of the ANUBIS project~\cite{Shah:2024fpl}. 
This encompasses the design and pre-construction phase of the detector at the Cavendish Laboratory and the CERN BB5 laboratory in 2022, as well as the construction and on-surface commissioning at BB5 in 2023.
After a brief overview of the Resistive Plate Chamber (RPC) detector technology, the construction and performance studies of RPCs along with the on-detector Front-End (FE) electronics are presented.

%%%%%%%%%%%%%%%
\section {The ANUBIS prototype: \proanubis}
\subsection {Design}
\label{sec:design}
\begin{figure}[ht]
    \centering
    \begin{subfigure}[b]{\linewidth}
        \centering
        \includegraphics[width=5.5 cm]{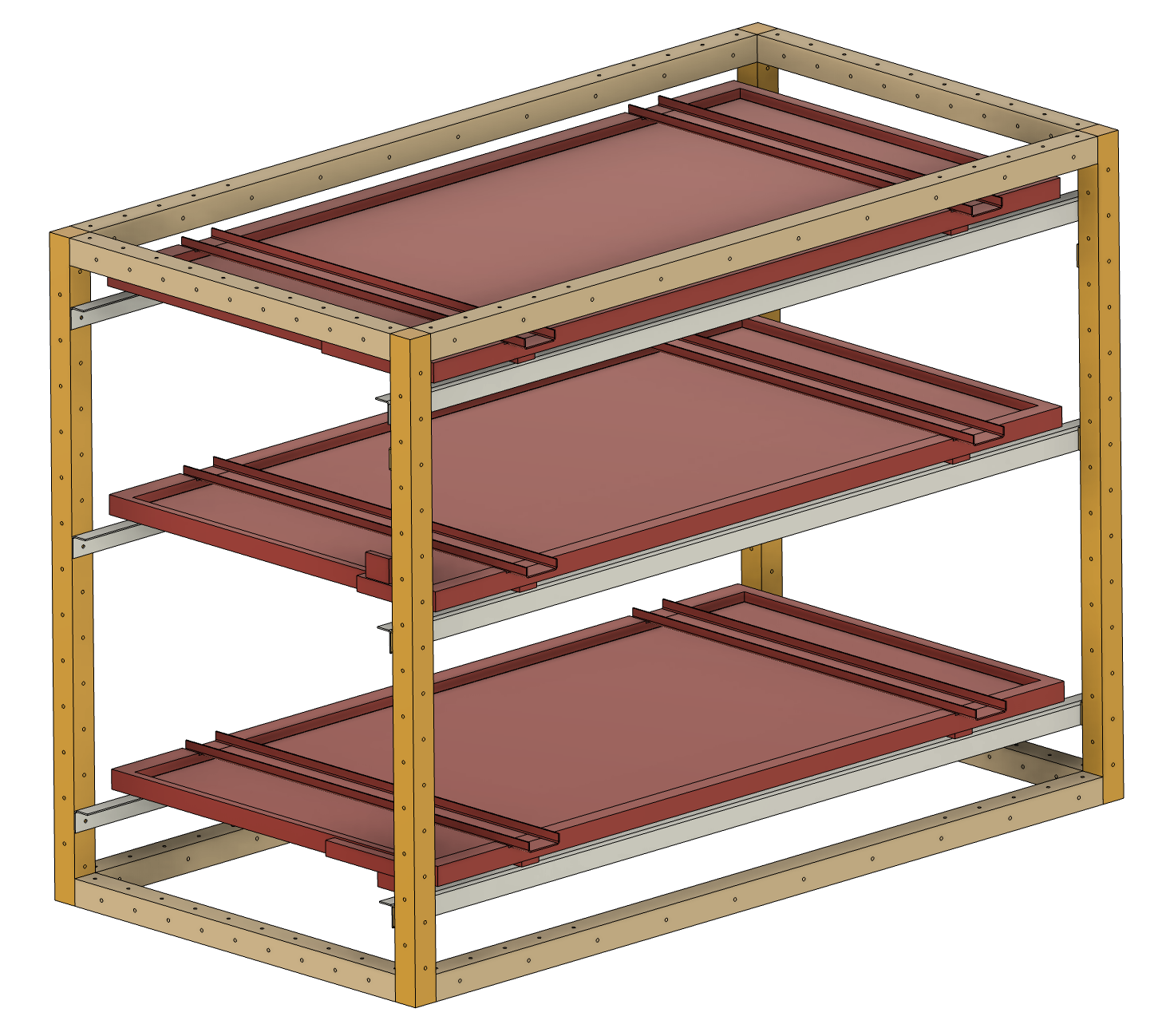}
        \caption{}
        \label{fig:proANUBIS_design_Top}
    \end{subfigure}
    \begin{subfigure}[b]{\linewidth}
        \centering
        \includegraphics[width=5.5 cm]{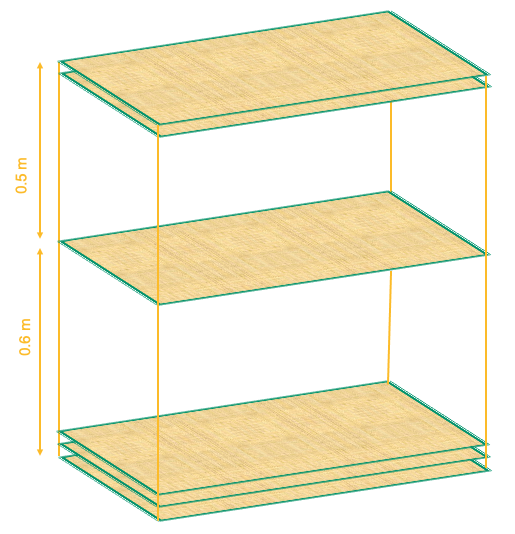}
        \caption{}
        \label{fig:proANUBIS_design_Low}
    \end{subfigure}
    \caption{
(a) The design of the \proanubis detector, showcasing the positioning of the three integrated tracking layers. 
(b) The arrangement of the tracker components within the \proanubis setup. 
At the bottom, three RPCs (a triplet) are shown, followed by a singlet in the middle, and a doublet on top.}
    \label{fig:proANUBIS_design}
\end{figure}

The configuration and geometry of the \proanubis detector are depicted in Figure~\ref{fig:proANUBIS_design}. 
The setup is designed to detect tracks of charged particles and reconstruct the vertices formed by these tracks. 
The \proanubis detector features three tracking layers, each measuring an area of $99~\cm~ \times 182~\cm$. The short side is conventionally referred to as the $\eta$ side, and the longer one the $\phi$ side\footnote{This naming convention is adopted from the ATLAS BIS7 project where the RPCs measure the $\eta$ and $\phi$ coordinates of a hit.}.
The bottom tracking layer, the triplet, comprises three individual RPC detectors, the top tracking layer contains a doublet of RPC detectors, and the middle layer is one single RPC detector, known as the singlet, as illustrated in this figure.
The use of multiple RPCs per tracking layer is driven by the requirement of efficiently detecting hits resulting from charged particles from $pp$ LHC collisions and cosmic rays. 
The top and middle (middle and bottom) tracking layers are separated by a vertical distance of 0.5~\metre~(0.6~\metre).
\enlargethispage{5mm}

To test the impact of the singlet chamber on the pattern recognition capabilities of \proanubis, the top RPC tracking layer was configured as a doublet instead of a triplet since six RPC detectors were produced. 
While the top and bottom layers alone would suffice for muon tracking, the singlet layer provides an additional hit, which reduces ambiguities when several charged particle tracks are detected in a given event, particularly for vertexing. 
The spacing between the tracking layers is optimised for maximum particle tracking accuracy within the maximum available detector volume. 
The overall geometry of the \proanubis detector is engineered to strike a balance between performance and practicality, considering the constraints imposed by the ATLAS underground cavern environment, \eg the limited space where \proanubis could be placed. 
These tracking layers feature front-end electronics developed for the ATLAS Phase I upgrade of sector 7 in the Inner Barrel Short section~(BIS7)~\cite{Pizzimento:2019slz,CERN-LHCC-2017-017} for signal collection, amplification, and processing, and interfacing to the data acquisition (DAQ) system.

\subsection {Detector construction}
The \proanubis detector is built upon recent RPC technology at ATLAS, specifically utilising detectors designed for the BIS7 upgrade~\cite{Massa:2020hjw,Pizzimento_2020}. 
These RPCs serve as a precursor for the upcoming Phase-II Barrel Inner (BI) RPC upgrade~\cite{CERN-LHCC-2017-020}. 
Noteworthy features of the BIS7 detectors include thinner Bakelite electrodes (1.4~\mm instead of the standard 1.8~\mm) with a graphite coating to improve rate capability, and a reduced gas gap width of 1~\mm, a departure from the conventional 2~\mm gap used in the legacy RPCs of the ATLAS muon system to improve the timing precision~\cite{Pizzimento:2024ndt}. 
These modifications result in a reduction in detector weight and size while improving detector performance. 
Equipped with copper readout strips and upgraded on-detector FE electronics, these RPCs demonstrate enhanced performance metrics such as improved time resolution, charge distribution, and signal collection efficiency. 
Furthermore, the BIS7 technology operates at nearly half the applied high voltage (HV) compared to ATLAS Phase-I RPCs for a given gas mixture, while maintaining a similar electric field. 
This characteristic reduces discharge probability and ensures operational stability.

The BIS7 technology was selected for \proanubis considering its compatibility with ANUBIS' performance requirements as outlined in the original proposal~\cite{Bauer:2019vqk} combined with its low weight and cost-effectiveness, which are important factors given the instrumented area of ANUBIS of about 1,700~m$^2$. 
Additionally, the technology offers commendable timing and spatial resolution, $\sim400$~\ps and a few \mm, respectively, making it an optimal choice for resolving high-multiplicity LLP decays and rejecting out-of-time tracks from cosmic ray particles or collision backgrounds~\cite{bib:anubis_v2}.

\subsection {Data acquisition system}
\begin{figure}[ht]
\centering
\includegraphics[width=6.0cm]{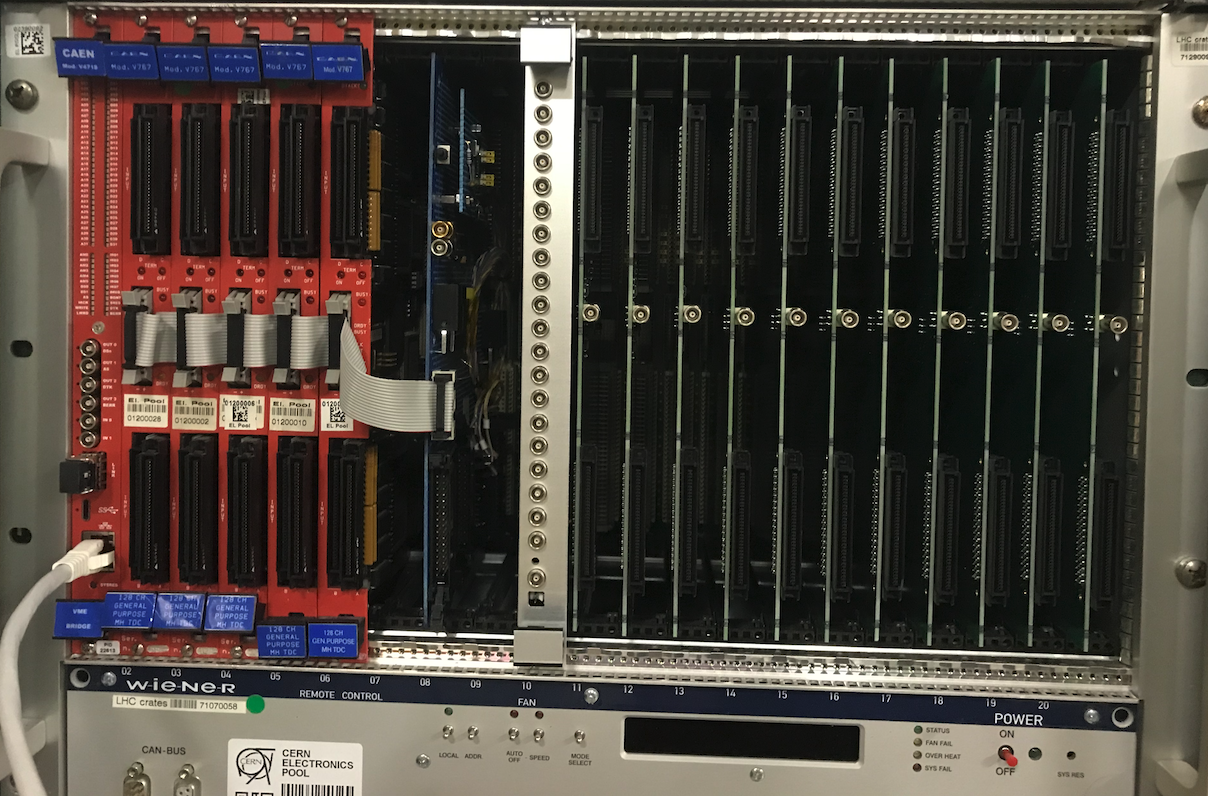}
\caption{
\label{fig:proANUBIS_DAQ_system}
A 6U VME crate with \proanubis DAQ components from left to right: the CAEN V4718 Ethernet controller card, six CAEN V767 TDCs, a signal translator board, the trigger logic board, and eleven trigger boards that are referred to as ``OR boards''.
} 
\end{figure}

\begin{figure}[ht]
\centering
\includegraphics[width=\linewidth]{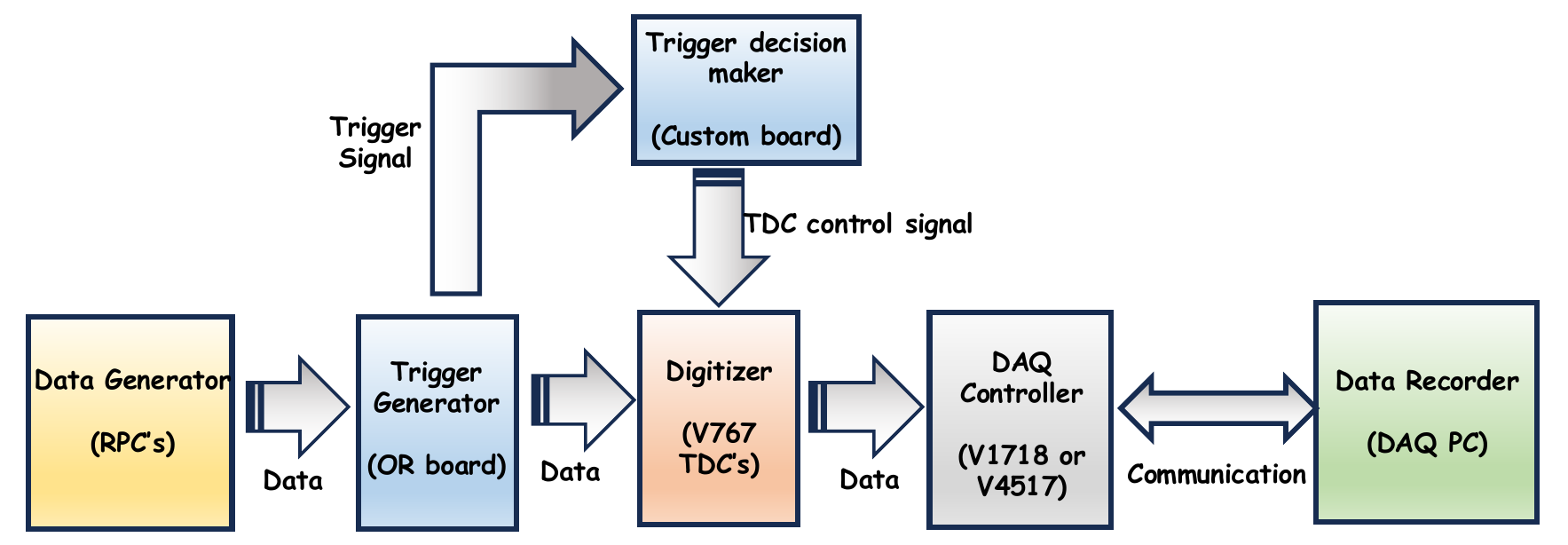}
\caption{
\label{fig:proANUBIS_DAQ_system_Diagram}
The data flow diagram of the \proanubis DAQ system up to the data recording stage.
} 
\end{figure}

The primary DAQ system employed in \proanubis is presented in Figures~\ref{fig:proANUBIS_DAQ_system} and~\ref{fig:proANUBIS_DAQ_system_Diagram} and comprises a combination of commercially available and custom-designed components. 
The heart of the DAQ system is a set of commercial CAEN V767 Time to Digital Converters (TDCs) featuring an 800~ps time resolution, along with a controller card -- either the USB-based CAEN V1718 model or the ethernet-based CAEN V4718 model. 
The controller serves as an interface to the data-taking server by facilitating communication and data transfer from the TDCs. 
During the 2024-25 Year End Technical Stop, the TDCs were upgraded to CAEN v1190A TDCs with an improved time resolution of down to 100~ps.

The trigger system of \proanubis is implemented through a combination of custom-made VME 6U electronics. 
The signal from the $\eta$ side of the detector is first received by trigger boards, also known as ``OR boards''. These OR boards generate a trigger signal by performing a logical OR of signals received from 32 input channels per OR board in a given 25~ns time window. 
The OR board then feeds the detector signal in the Low-Voltage Differential Signaling (LVDS) format through to the TDCs, and provides the trigger signal in TTL format to the trigger logic board. 
The trigger logic board performs the ultimate trigger decision using a simple $N$ out of $M$ logic, where $M=6$ and $N$ is typically 4 to select coincident hits in several detector layers when a charged particle passes through the detector.
The output of the trigger logic board is then passed to a signal translator, which combines it with the 40~kHz clock signal of the LHC.  
The output of the $\phi$ side of the detector is fed directly through to the TDCs, bypassing the trigger. For more details of the \proanubis DAQ system and signal processing, see the forthcoming \proanubis commissioning report.

The remainder of this manuscript will focus on the construction of the RPCs, and the test procedures carried out to ensure they are of sufficient quality for the operation of \proanubis.

\section {RPC construction and performance} 
\label{Sec:RPC_cons_perform}

\subsection {Gas gap quality assurance \& quality control procedures}  
\label{Sec:gas_gas_QC}
\begin{figure}[ht]
\centering
\includegraphics[width=5.4cm, height=5.4cm]{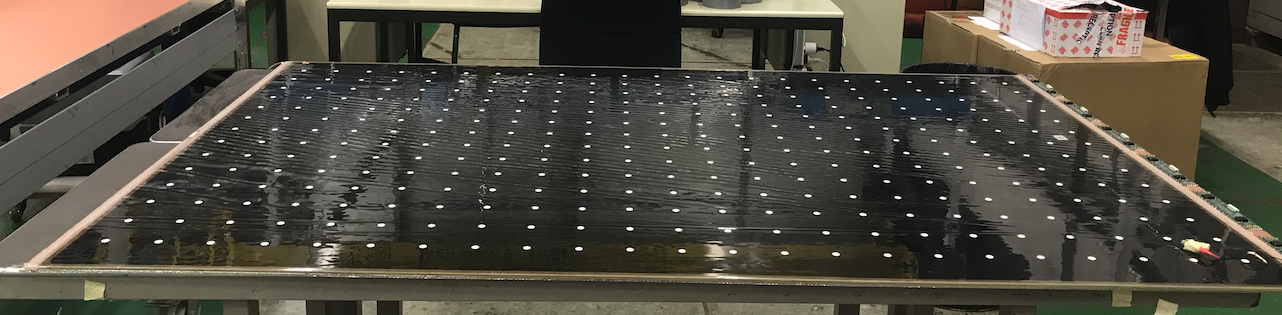}
\caption{
One of the gas gaps upon reception at CERN. 
The white patches on the gas gap mark the positions of the spacers.} 
\label{fig:RPCspacer}
\end{figure}

The BIS7 RPC gas gaps, as shown in Figure~\ref{fig:RPCspacer}, undergo several Quality Assurance and Quality Control (QA/QC) measures. Initially, these gaps experience an acceptance testing process at General Tecnica Engineering (GTE), the commercial manufacturer. 
This process entails a thorough visual inspection of the Bakelite plates utilised in gap construction to detect any surface defects. 
Furthermore, measurements are conducted to assess the surface resistivity of the graphite-coated Bakelite electrodes, targeting a value of ${\sim500~\Omega/\square}$, and to ensure its uniformity. 
The gas gap spacer thickness, which is  1~\mm, is also confirmed with a tolerance of 15~\mum~\cite{Massa:2020hjw}.

\begin{figure}[ht]
\centering
\includegraphics[width=8 cm, height=5.4cm]{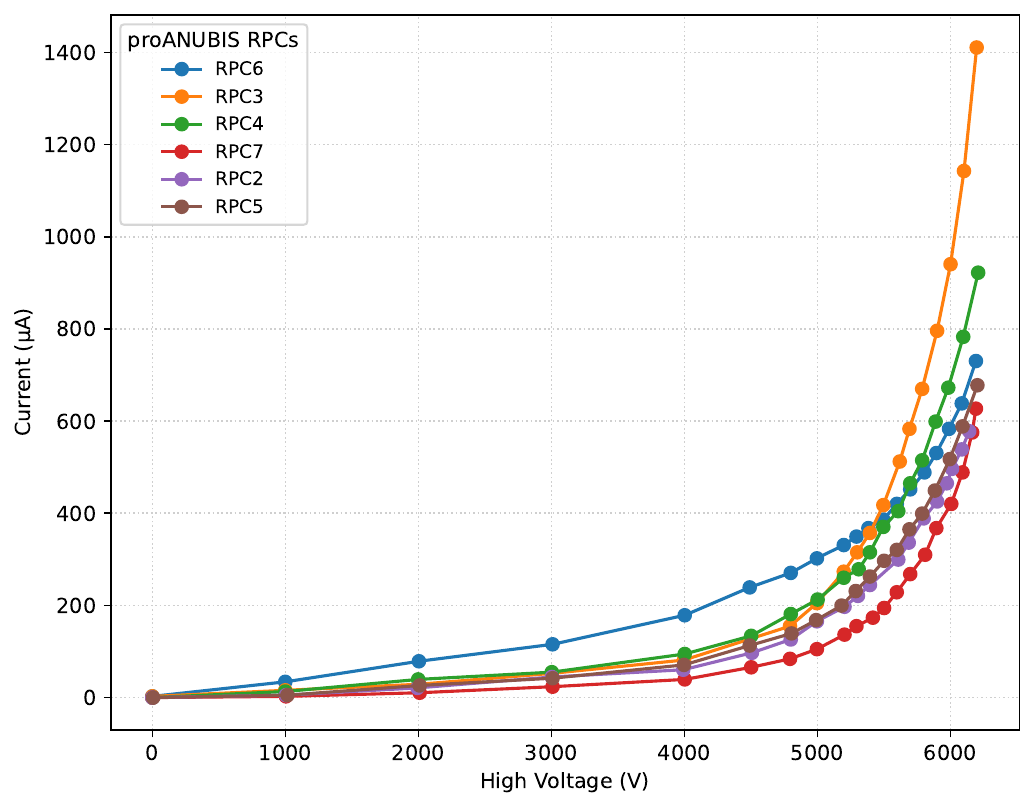}
\caption{
The IV characteristics for the \proanubis gas gaps measured at CERN before the detector assembly. 
Each RPC is uniquely identified by a tag, \eg RPC2.} 
\label{fig:Gas_gaps_IV}
\end{figure}

Upon assembly, the polymerisation of the oil coating the inner surface of the resistive plate undergoes scrutiny through a destructive test. 
This involves opening small representative samples of gas gaps, constructed in parallel, for visual inspection. 
Proper glueing between the external Polyethylene Terephthalate (PET) insulator and the electrodes is also verified through visual inspections, ensuring the absence of air bubbles. 
Subsequently, gas gaps undergo tests to evaluate gas tightness and HV insulation.
Finally, their Volt-Amperometric (IV) characteristics are determined by operating them with the standard RPC gas mixture and measuring the current $I$ across the gas gap as a function of voltage $V$ of up to 6.3 kV, see Figure~\ref{fig:Gas_gaps_IV}. 
This test verifies that the ohmic leakage current, which is linear as a function of $V$, is small, and that the transition to avalanche mode, characterised by the exponential rise of the current, occurs as expected.
Following this stage, the gaps are shipped to CERN for further testing, where some of these tests are repeated, \ie the visual inspections, the gas tightness, and the IV tests.

In certain cases, additional tests for long-term stability are conducted on gas gaps at CERN, which include subjecting them to one week of gamma radiation exposure (these tests are done mostly on a subset of gas gaps for the ATLAS BIS7 project). 
During irradiation, a current across the gas gaps is induced due to avalanches caused by random interactions of the incoming photon flux with the gas. 
Hence, the gas gaps undergo conditioning throughout the irradiation period, gradually increasing the $V$ applied and hence the current across the gap at a controlled rate. 
This conditioning process enhances gas gap performance by eliminating small defects on their surface and effectively reducing both linear, \ie ohmic, and exponential parts of the IV curves. 
Any gaps that fail to exhibit this improvement are discarded.

For the \proanubis detector, seven gas gaps underwent testing (labelled as RPC1-RPC7) and all met the stringent QA/QC criteria. 
Of these, only six were used in \proanubis, with the last reserved as a spare and for stand-alone performance tests.

\subsection {Front-End boards QA/QC procedures}  \label{Sec:FE_board_testing}
\begin{figure}[ht]
\centering
\includegraphics[width=8.5cm]{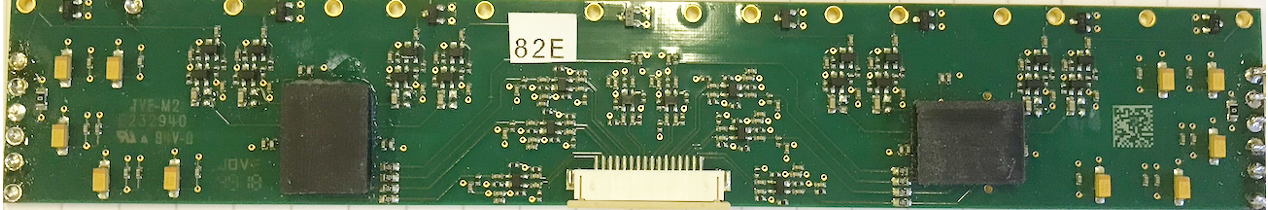}
\caption{A picture of the BIS7 type FE board used in \proanubis that handles eight readout channels.} \label{fig:FE_actual_board}
\end{figure}

The FE electronics board designed for the readout of the new generation of BIS7 RPCs with a 1~mm gas gap was developed by the ATLAS muon group~\cite{Pizzimento:2019slz}. 
Figure~\ref{fig:FE_actual_board} shows one of these readout Printed Circuit Boards (PCBs). The board integrates a pre-amplifier and a discriminator in SiGe BiCMOS (Silicon Germanium Bipolar Complementary Metal-Oxide-Semiconductor) technology, as well as some other minor modifications that result in enhanced performance of the RPCs, such as a one order of magnitude higher rate capability and improved space-time resolution, featuring a reduced power consumption overall~\cite{RCardarelli_2013, Pizzimento:2019slz,Pizzimento:2024ndt}. 
Notable features of this new FE board are summarised in Table~\ref{tab:FE_comp}.

\begin{table}[ht]
\begin{center}
\caption{Comparison of some of the main parameters of new versus old FE boards~\cite{Pizzimento:2024ndt}.}
\label{tab:FE_comp}
\vspace{0.1cm}
\begin{tabular}{l|l|r}
\hline
\textbf{Parameter} & \textbf{Standard RPC} & \textbf{BIS7 RPC}\\
\hline
\multicolumn{1}{l|}{Effective threshold} & 2-3~mV & 0.2-0.3~mV\\
\multicolumn{1}{l|}{Power consumption} & 30~mW/ch & 15~mW/ch\\
\multicolumn{1}{l|}{Technology} & GaAs & BJT Si + SiGe\\
\multicolumn{1}{l|}{Discriminator} & Embedded & Separated\\
\multicolumn{1}{l|}{TDC embedded} & No & No\\
\hline
\end{tabular}
\end{center}
\end{table}

Each FE board has eight pre-amplifier channels that are fed into two custom ASIC discriminators, each processing four channels. Additionally, the boards are equipped with integrated LVDS drivers. 
As avalanche electrons generated by charged particles traverse the RPC gas gap, drifting towards the positive electrode, electromagnetic signals are induced in the readout strips on either side of the gas gap that are processed by the FE boards. 
These signals have opposite polarity since they are attached at opposite sides of the gas gap.
Specifically, \textit{eta} ($\eta$) and \textit{phi} ($\phi$) boards receive negative and positive signals, respectively, from the same event. 
However, the ASIC discriminator requires a negative signal input, requiring two versions of the readout boards: \(\eta\) and \(\phi\). 
A \(\phi\) board requires a signal inverter before the discriminator to convert the positive input into a negative one, but an \(\eta\) board does not. 
In practice, this was controlled on the board itself by soldering a capacitor on dedicated positions. 
In total, 12 FE boards are required per singlet, with eight for the 182~cm long \(\phi\) side and four for the 99~cm long \(\eta\) side.

At CERN, the FE boards are subjected to testing procedures to ensure operational functionality across all input channels. 
One approach focuses on the threshold voltage parameter of the discriminator, $V_{\rm TH}$. 
To assess the FE board performance, its eight input channels are perturbed utilising a metal spring probe connected to a signal generator, and output signals are acquired through a TDC. This is done for $V_{\rm TH}$ values varied within a defined range of $1.5\,\text{V} \le V_{\rm TH} \le 1.7\,\text{V}$~\cite{Massa:2020hjw}.

\begin{figure}[ht]
\centering
\begin{subfigure}[b]{\linewidth}
    \centering
    \includegraphics[width=7.8cm, height=4.3cm]{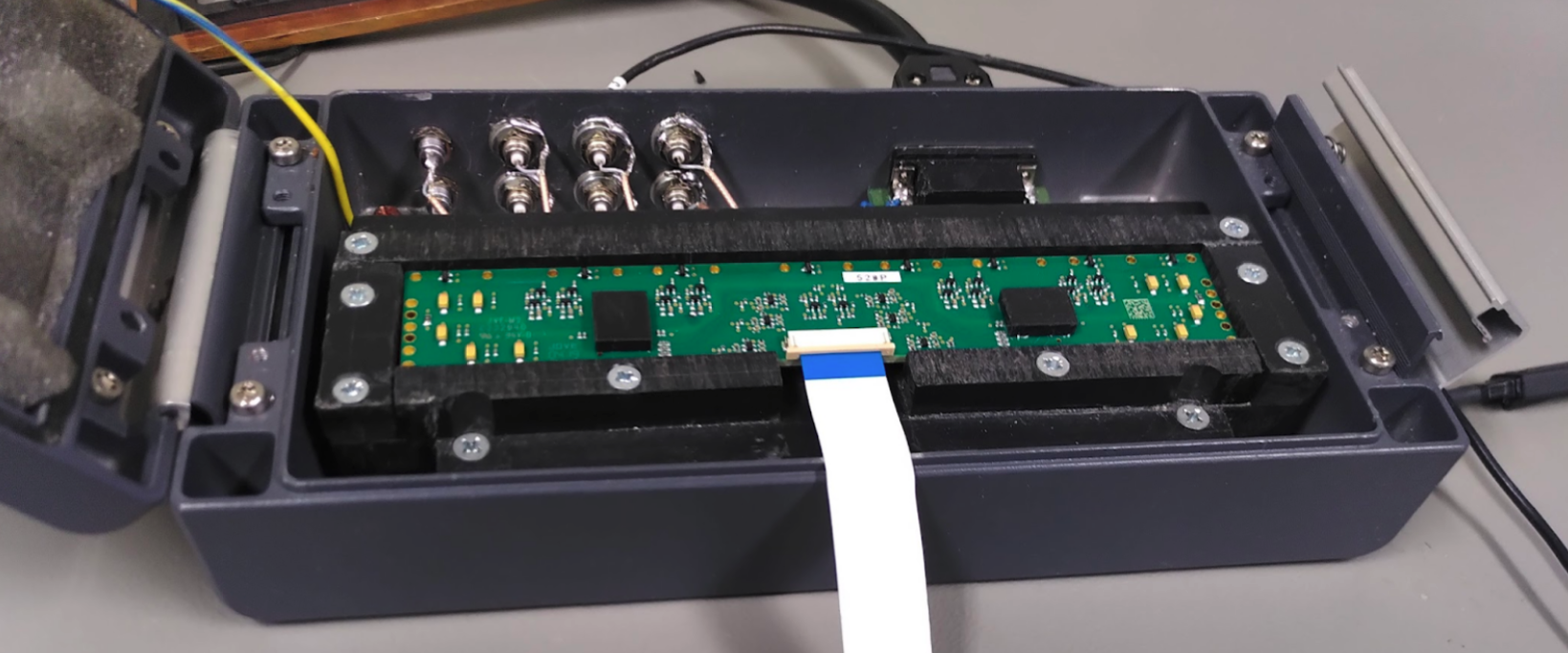}
    \caption{}
    \label{fig:FE_board_testing_Top}
\end{subfigure}
\begin{subfigure}[b]{\linewidth}
    \centering
    \includegraphics[width=7.8cm, height=4.3cm]{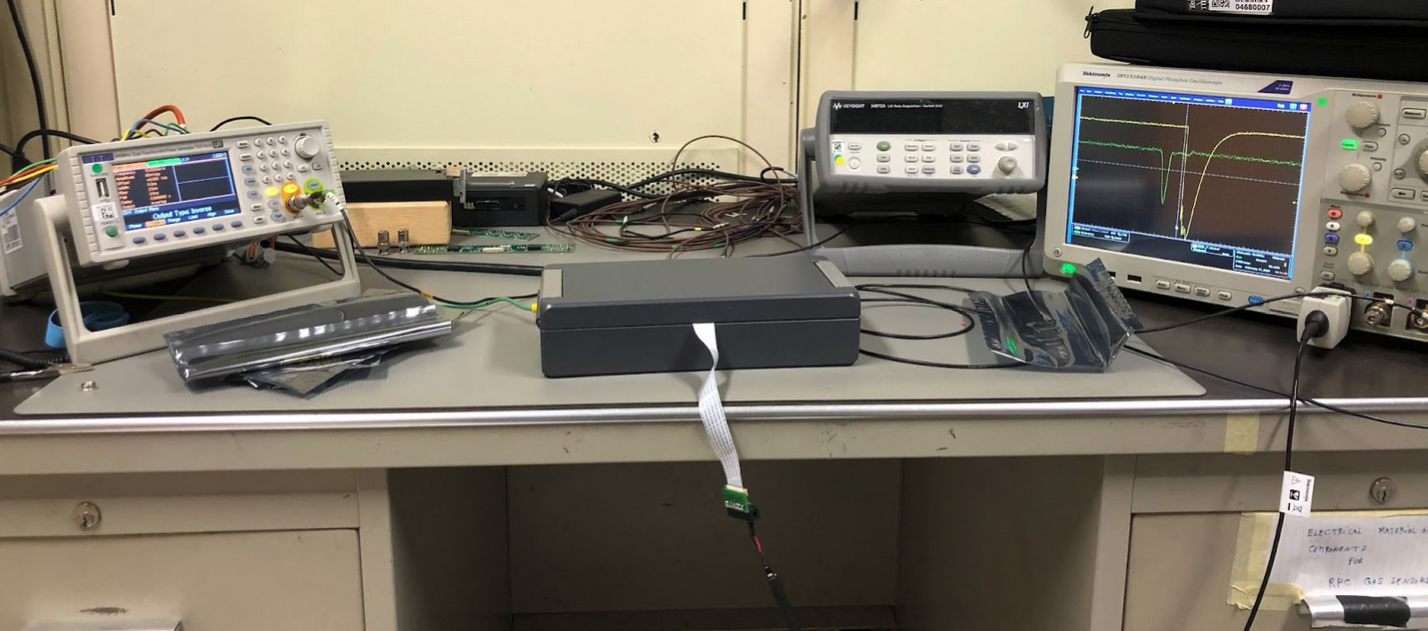}
    \caption{}
    \label{fig:FE_board_testing_Low}
\end{subfigure}
\caption{(a) The FE board inside a test box at CERN. (b) The full test setup with a closed test box to avoid any electromagnetic noise interference, a signal generator being used to inject test signals at the input of the FE board via flat ribbon cable, and the GHz resolution oscilloscope used to display output from the FE boards.} 
\label{fig:FE_board_testing}
\end{figure}

In addition to $V_{\rm TH}$ testing, a pulse generator is used to inject signals of a few nanoseconds as input to a multiplexer. This setup allows the pulse to be input to channels 1 to 8 for assessment. The multiplexer expedites the process and refines the workflow for testing batches of boards~\cite{Swallow:2883099}. The test-box, as depicted in Figure~\ref{fig:FE_board_testing}, is specifically designed for testing the BIS7 readout boards. Signals for the input channels are injected via eight ports at the back of the box, and the injected pulse and the board’s output are compared using an oscilloscope with a resolution of {$\ge$ 1~GHz}. 
The testing procedure entails injecting a signal into one input channel and subsequently measuring each output channel, with four possible outcomes considered:
\begin{enumerate}
 \item A signal peak observed in the corresponding output channel, indicating expected behaviour.
 \item No signal detected in the corresponding output channel, signalling a `dead' channel.
 \item Signal observed in an unexpected channel, typically adjacent to the injection channel, usually referred to as `cross-talk'.
 \item Unusual behaviour, \eg truncated signals, is observed in the output channel.
\end{enumerate}
A total of 72 FE boards were tested. 
Each of them showed the first outcome, hence meeting our QC requirements of the FE boards to be accepted for installation. 
This provided the required number of boards for the assembly of RPCs for the \proanubis detector.

\subsection {Strip panel assembly}
\begin{figure}[ht]
\centering
\includegraphics[width=7.7 cm, height=4.7cm]{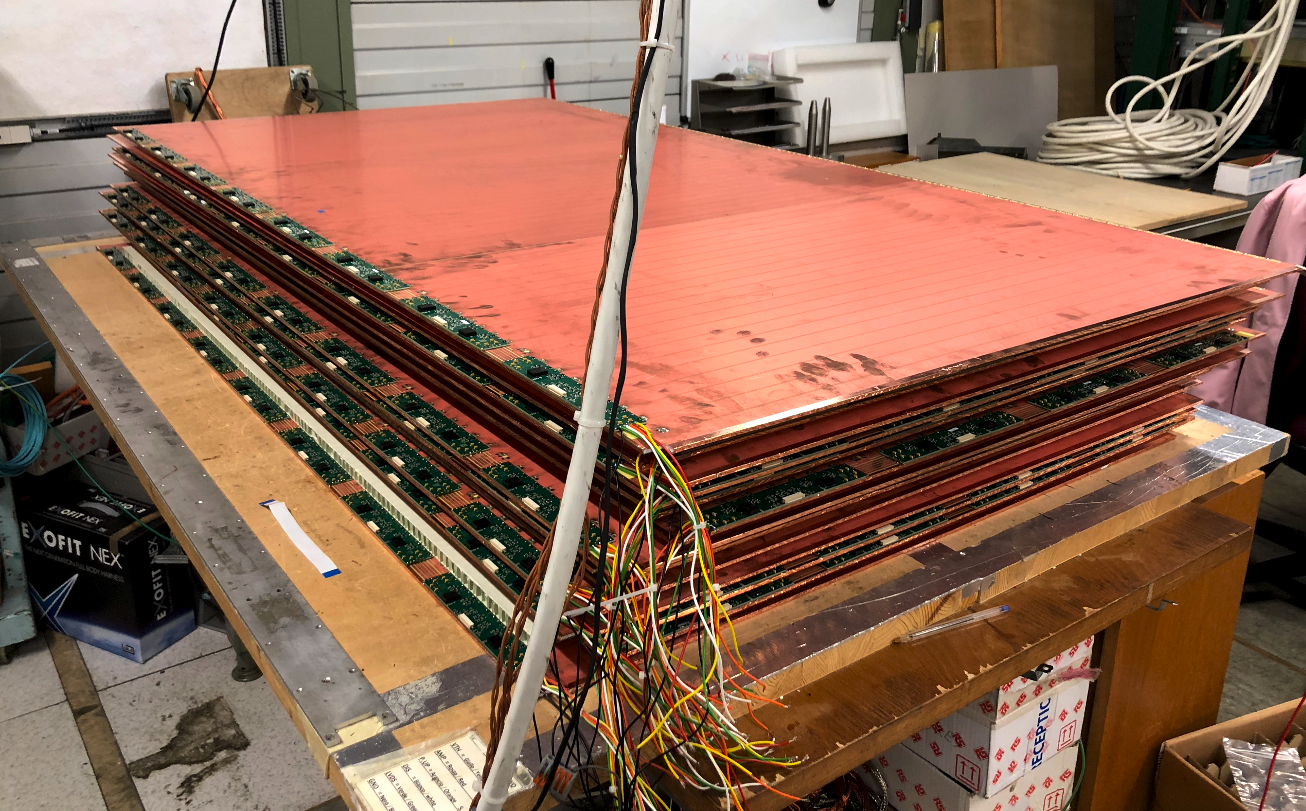}
\caption{A stack of BIS7 RPC strip panels after having their termination resistors and FE boards installed.} 
\label{fig:Strip_panels}
\end{figure}

Prefabricated copper readout strip panels with an approximate pitch of 25~mm were modified and assembled to detector requirements, as shown in Figure~\ref{fig:Strip_panels}. 
Each panel measures 182~\cm $\times$ 99~\cm and two such panels, with strips running in perpendicular directions, compose one RPC singlet. 
Each singlet has one panel with 64 strips running parallel to the short (99~\cm) dimension referred to as the \(\phi\)-panel, the other is the \(\eta\)-panel which has 32 strips running parallel to the long (182~\cm) dimension, along with some non-active area on each panel for the readout electronics. 
For each panel the strips were affixed to a Forex sheet that acts as an insulation layer to separate the strips from a copper ground plate.
These panels were then prepared by soldering termination resistors to one end of each strip to match the impedance of the FE electronics and eliminate signal reflection, and the FE boards described in Section~\ref{Sec:FE_board_testing} to the other end of each strip to read out the signal. 
After preparation, these panels were tested by evaluating their FE boards’ ability to read a test signal induced by touching one strip at a time with a finger and observing the signal on an oscilloscope. 
This ensures the absence of any major defects in the panels or their electronics. 
This test showed that more than 99.5\% of the readout strips successfully read out a signal.

\subsection {RPC singlet assembly}
The assembly parts for the RPC production, including gas volumes, copper strip panels, FE boards, and mechanical frames, were collaboratively sourced from Italy and Germany. 
Following production, these materials were transported to CERN for assembly and subsequent testing through a series of QA/QC procedures, as detailed in Section~\ref{Sec:gas_gas_QC}.
Upon arrival at CERN, the gas gaps underwent further testing, while the copper strip panels were affixed to Forex insulators to form the readout panels. 
Additionally, the FE boards underwent the testing procedures outlined in Section~\ref{Sec:FE_board_testing}.

\begin{figure}[ht]
\centering
\includegraphics[width=4 cm, height=5.5cm]{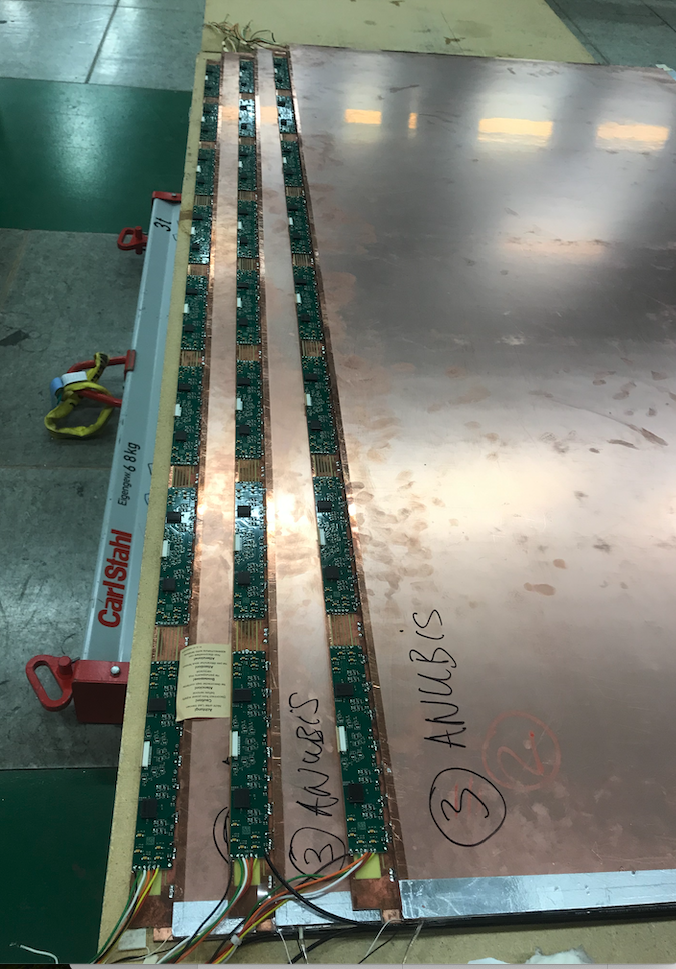}
\caption{Three RPC singlets stacked on top of each other following assembly.}
\label{fig:RPC_Stack}
\end{figure}

The quality of the gas gaps was verified through measurements of key parameters such as ohmic current and the knee point in the IV characteristic. 
The ohmic current refers to the linear ($V\propto I$) part of the IV characteristic, \ie the behaviour up to $V\lesssim4$~kV in Figure~\ref{fig:Gas_gaps_IV}), and the knee point refers to the voltage when this linear behaviour `breaks down' and the exponential growth of $I$ with $V$ begins, \ie $V\approx 4$~kV for Figure~\ref{fig:Gas_gaps_IV}. 
The gas gaps are then sandwiched between readout panels to create RPC singlets.
Figure~\ref{fig:RPC_Stack} shows the arrangement of RPC singlets in a triplet stacked on top of each other after assembly. 
These singlets are then tested using cosmic rays, as described in Section~\ref{Sec:RPC_Cosmic_testing}. 
Additionally, the FE board quality is tested by measuring the random counting rate and the current drain. 
Subsequently, the singlets are integrated into mechanical frames to form a triplet, a singlet, and a doublet as indicated in Figure~\ref{fig:proANUBIS_design} and discussed in more detail in Section~\ref{sec:integration}.

\begin{figure}[ht]
\centering
\includegraphics[width=7.3 cm, height=5.2cm]{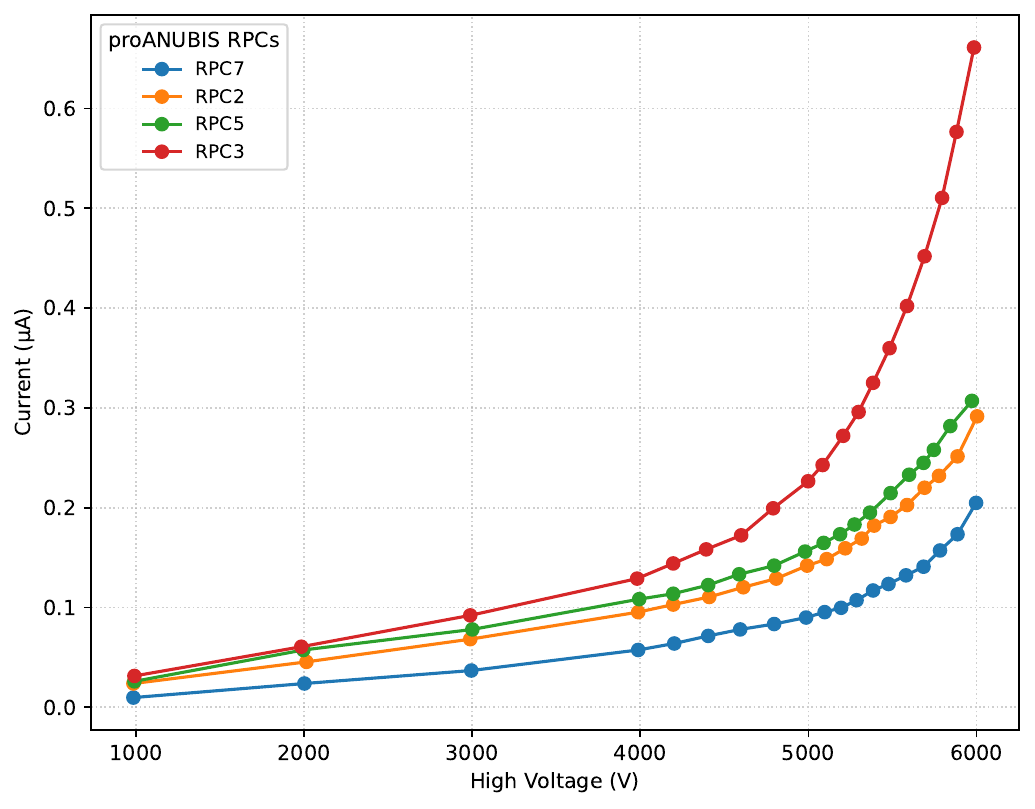}
\caption{
The IV characteristics for the proANUBIS gas gaps after
the detector assembly.
}
\label{fig:RPC_Stack_IV}
\end{figure}

Figure~\ref{fig:RPC_Stack_IV} presents IV measurements of four representative gas gaps. 
In the region where the detector is expected to be operated, $I<1~\mu{\rm A}$ was measured, demonstrating the expected behaviour of these detectors. 
These measurements served as a performance indicator for the reliable operation of the RPC detectors, validating their use for the \proanubis project.

\subsection {Performance studies using cosmic ray muons} 
\label{Sec:RPC_Cosmic_testing}
\begin{figure}[ht]
\centering
\includegraphics[width=7.5 cm, height=4.8cm]{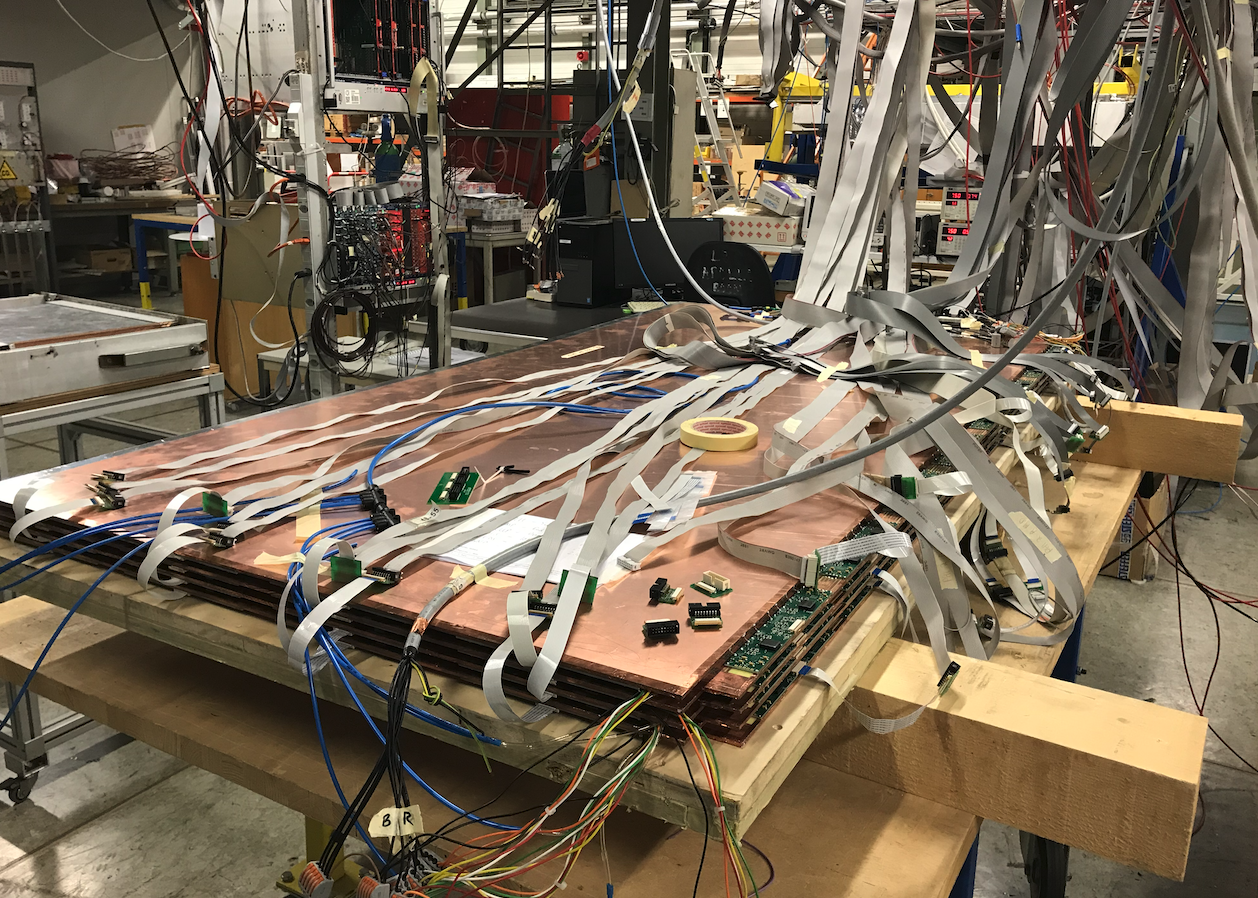}
\caption{The RPC configuration employed at the CERN BB5 lab for efficiency measurements. The experimental arrangement consists of three RPCs stacked on top of each other and connected to FELIX DAQ system to evaluate their performance.} \label{fig:RPC_setup}
\end{figure}

To evaluate the efficiency of each RPC, post-assembly measurements were conducted at the CERN BB5 laboratory with cosmic ray muons and using a standard gas mixture, as depicted in Figure~\ref{fig:RPC_setup}. This standard gas mixture comprises 94.7$\%$ C$_{2}$H$_{2}$F$_{4}$ (Freon, commercially known as R134a), 5$\%$ i-C$_{4}$H$_{10}$ (isobutane), and 0.3$\%$ SF$_{6}$~\cite{Su:2025gcf}. 
Freon acts as the main ionisation gas, while isobutane acts as a quencher to limit the formation of secondary avalanches by absorbing photons generated from recombination processes. 
SF$_{6}$ is an electronegative gas and is used to capture free electrons, thereby suppressing the onset of streamers, a transient discharge from the surface of the electrodes, by limiting free charges within the gas volume.

\begin{figure}[ht]
\centering
\begin{subfigure}[b]{\linewidth}
 \centering
\includegraphics[width=6.6 cm, height=5.1cm]{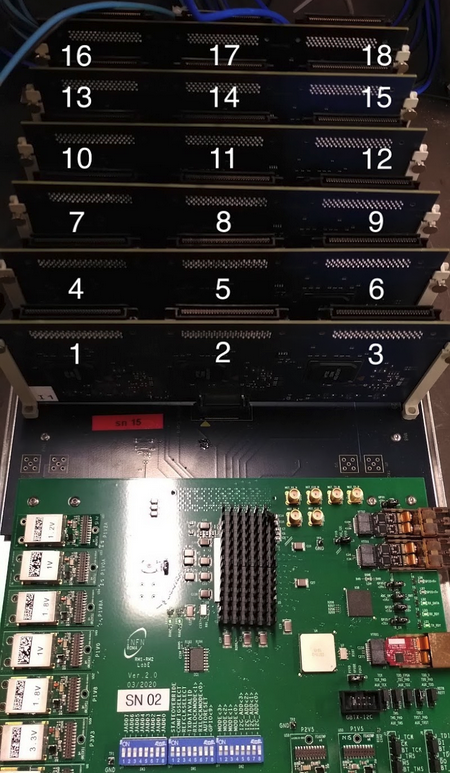}
\caption{}
\end{subfigure}
\begin{subfigure}[b]{\linewidth}
 \centering
\includegraphics[width=6.6 cm, height=5.1cm]{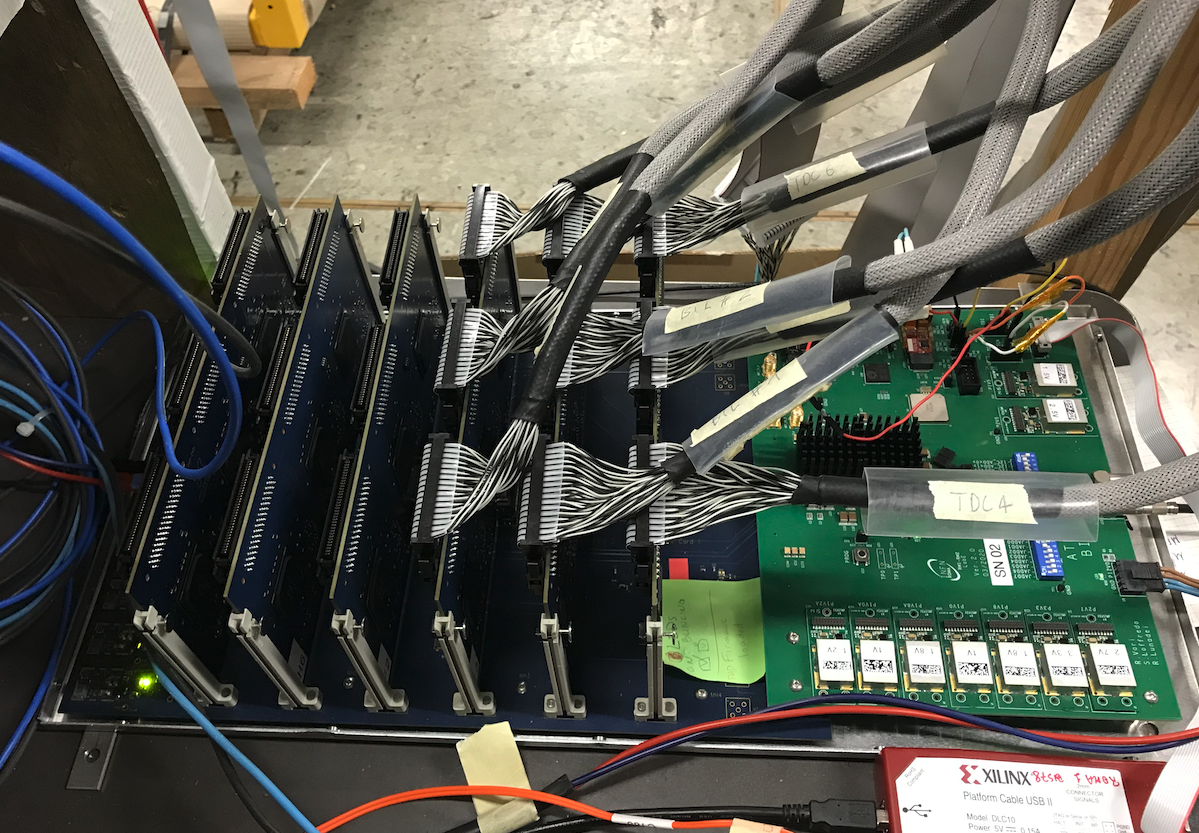}
\caption{}
\end{subfigure}
\caption{(a) The FELIX DAQ system showing different cards with TDC chips numbered from 1 to 18, a serialiser, an amplifier, a shaper, and a discriminator as well as the PCB BIS7 Pad board on the bench. (b) Three RPCs connected to three cards (9 TDC chips) of the FELIX DAQ setup for efficiency purposes.} 
\label{fig:flex_setup}
\end{figure}
%\hspace{1pt}

The measurement setup involved a coincidence setup featuring three RPCs stacked upon each other with nearly perfect alignment. 
Two of the RPCs, \eg top and bottom, are maintained at a fixed HV of $V=5.8$~kV, ensuring an expected efficiency of over 95$\%$. 
These RPCs provide trigger signals, while the third, \eg middle, RPC acts as a test chamber to measure its performance as a function of the applied HV and other parameters. 
Signals from the FE boards of BIS7 RPCs were digitised through High Performance TDCs (HP TDCs) with a time resolution of 200~\ps~\cite{Christiansen:1067476}. 
The hit data were collected by the BIS7 Pad~\cite{Ref_AMD_KINTEX}, an FPGA-based board employing a local trigger coincidence with a 2/3 majority logic to select muon candidates passing through the three RPC gas gaps. 
These selected events were then recorded by the FELIX (FrontEnd Link Interface eXchange~\cite{Anderson_2016}) data acquisition board, which was custom-developed for the ATLAS Phase-II upgrade. 
The experimental setup, inclusive of HP TDC, Pad, and FELIX boards, facilitated the reading of an entire BIS7 triplets, transmitting data to {\mbox{FELIX}} at 320~Mb/s, is shown in Figure~\ref{fig:flex_setup}.

\begin{figure}[ht]
\centering
\includegraphics[width=8.8 cm, height=6.0cm]{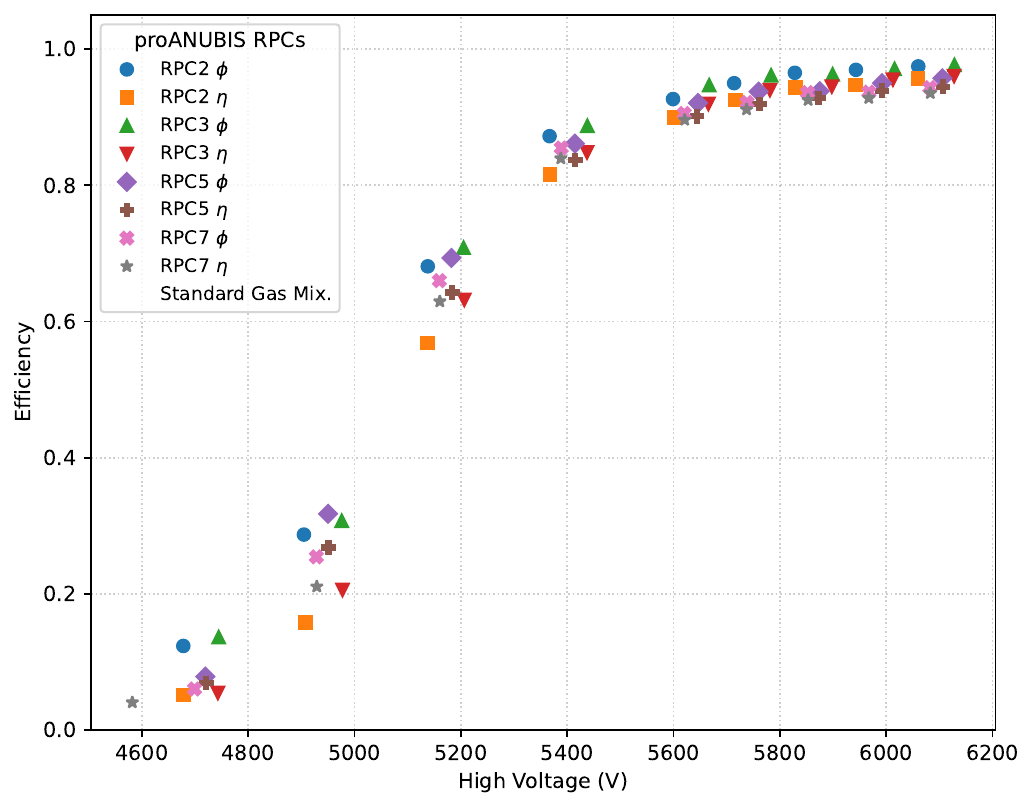}
\caption{Detector efficiency (\%) studies conducted for four singlet RPCs (for both \(\eta\) and \(\phi\) sides).} \label{fig:RPC_setup_Efficiencies}
\end{figure}

Events were recorded whenever a coincidence between the top and bottom RPCs occurred within a 100~\ns time window. 
Singlets were deemed acceptable if they exhibited an efficiency exceeding 95$\%$ in the plateau region of Figure~\ref{fig:RPC_setup_Efficiencies}, noise levels below 1 Hz/cm$^2$, fewer than 1$\%$ dead channels, and an average cluster size not exceeding 3 strips.
The experimental setup depicted in Figure~\ref{fig:RPC_setup_Efficiencies} yielded an estimated efficiency of over 95$\%$ in each case.

\subsection {RPC chamber integration}
\label{sec:integration}
Generally, the RPC module is composed of three BIS7 RPC singlets stacked directly on top of one another, housed in an aluminium mechanical frame. 
The frame encloses the RPCs with access available for readout through the different RPC planes both on \(\eta\) and \(\phi\) sides. 
A set of `skins’, \ie thin steel sheets, is put in place to cover the top and bottom of the RPC module, reinforced by four crossbars on both sides. 
A set of shims underneath those crossbars secures the RPCs in position within the module and counteracts the pressure from the gas gaps within each RPC in the triplet that could potentially warp them. 
These structures provide mechanical stability to the detectors, in particular in case of overpressure in the gas supply. 
Together with the steel `skins', they also act as a Faraday cage, shielding the detectors and sensitive FE electronics against external electromagnetic interference.  

\begin{figure}[ht]
\centering
 \begin{subfigure}[b]{0.49\linewidth}
  \centering
 \includegraphics[width=4.2 cm, height=4.2cm]{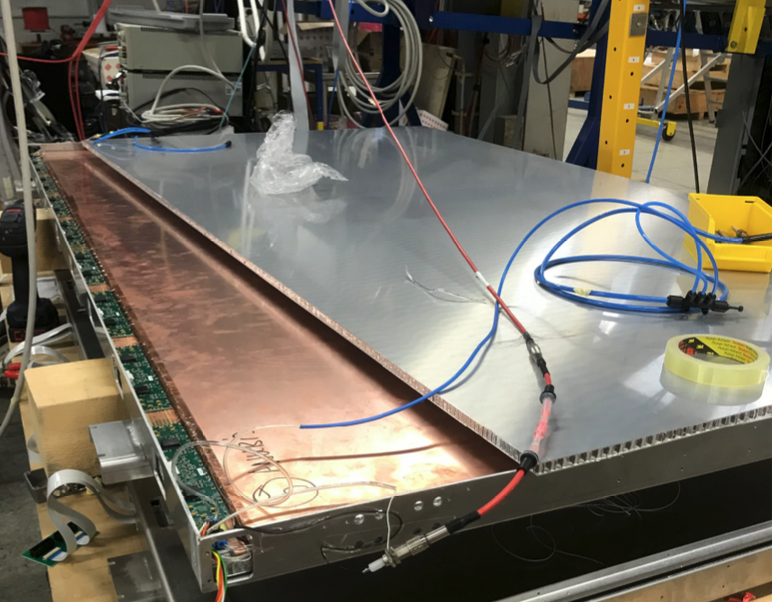}
 \caption{}
 \label{fig:RPC_full_integration_setup_A}
\end{subfigure}
 \begin{subfigure}[b]{0.49\linewidth}
  \centering
 \includegraphics[width=4.3 cm, height=4.2cm]{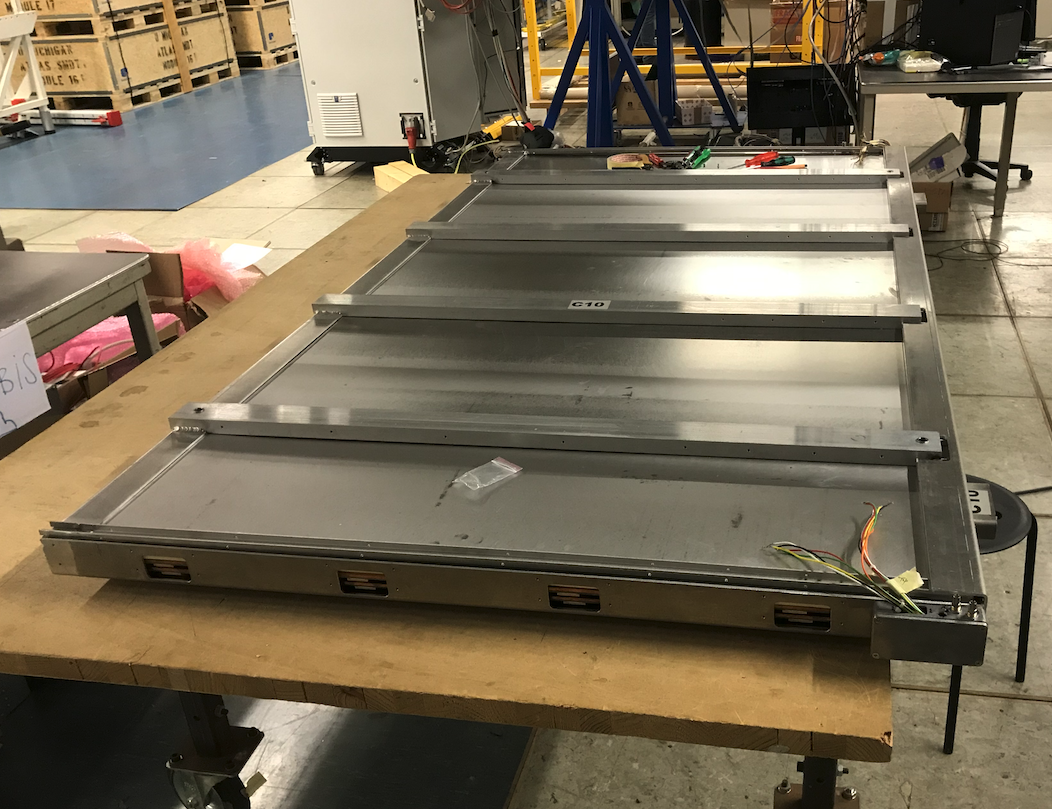}
 \caption{}
 \label{fig:RPC_full_integration_setup_B}
\end{subfigure}
 \begin{subfigure}[b]{0.49\linewidth}
  \centering
 \includegraphics[width=4.2cm, height=2.7cm]{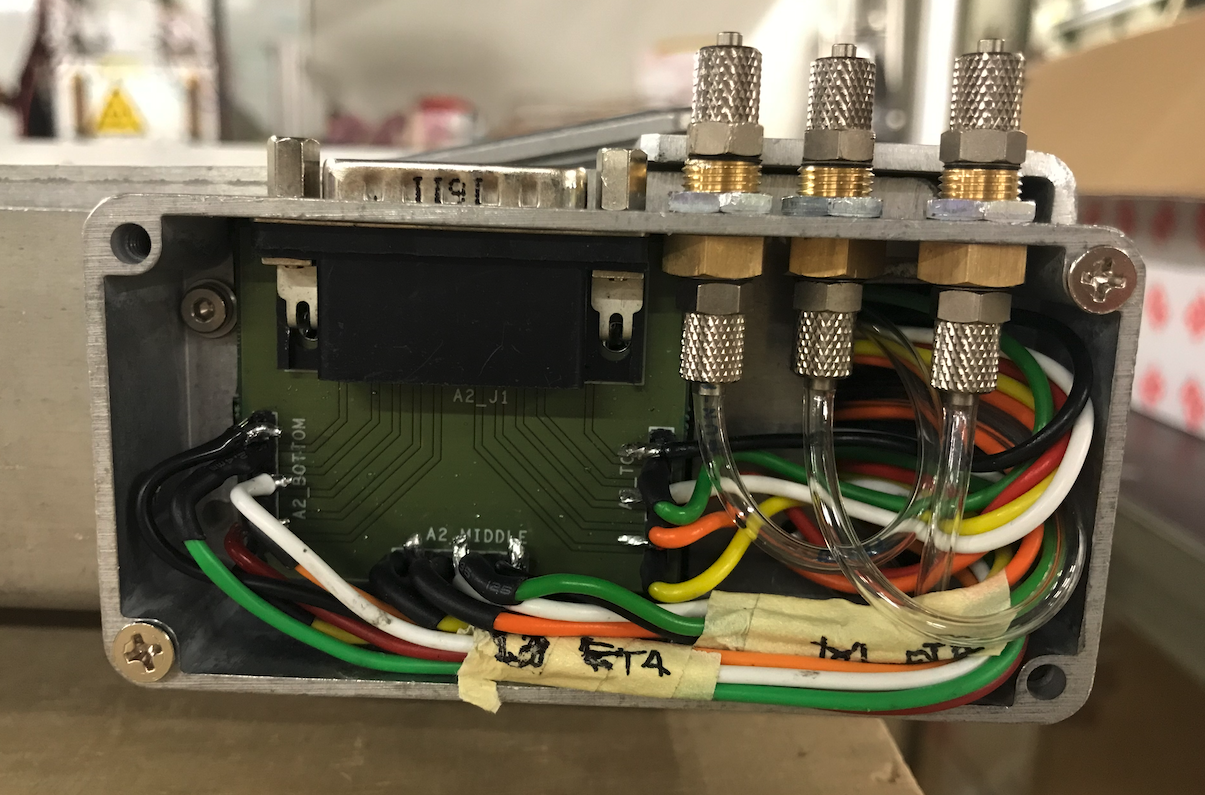}
 \caption{}
 \label{fig:RPC_full_integration_setup_C}
\end{subfigure}
 \begin{subfigure}[b]{0.49\linewidth}
  \centering
 \includegraphics[width=4.2cm, height=2.7cm]{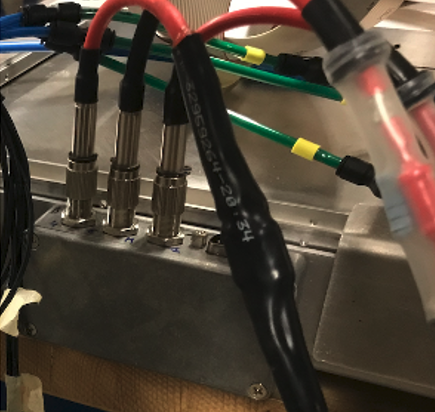}
 \caption{}
 \label{fig:RPC_full_integration_setup_D}
\end{subfigure}
 \caption{
 (a) A 12~\mm thick aluminium honeycomb structure is used as a dummy singlet to replace one in the frame and prevent any detector movement. 
 (b) An RPC module housing a pair of RPC detectors (a `doublet') within a metal frame to provide mechanical stability and act as a Faraday cage. 
 (c) An example of the gas integration setup box for a triplet RPC configuration, the tubing allows gas flow to the RPCs. 
 (d) HV cables connected via cable connectors to provide HV to individual RPCs.
 }
\label{fig:RPC_full_integration_setup}
\end{figure}

As described in Section~\ref{sec:design}, three arrangements of RPCs were used in \proanubis when integrating them within steel structures: a triplet, doublet, and a singlet. 
The integration process replaces empty spaces in the singlet and doublet structure cages with 12~\mm thick aluminium honeycomb sheets to prevent any detector movement, see Figure~\ref{fig:RPC_full_integration_setup_A}. 
The connections to gas, HV, and LV supplies were facilitated through rigid aluminium boxes, as shown in Figure~\ref{fig:RPC_full_integration_setup}. 
After full integration and assembly, the leakage current and efficiency measurements were re-evaluated to ensure the integrity of the detector’s performance, and no issues were observed.

\section {Commissioning of \proanubis at the CERN BB5 laboratory}
The commissioning process for \proanubis commenced at the CERN BB5 lab, where QA/QC procedures were conducted on the gas gaps, RPC panels, and FE boards. 
Upon successfully qualifying all individual components (as outlined in Section~\ref{Sec:RPC_cons_perform}), the RPC detectors were assembled, followed by the integration of RPC modules. 
This integration process involved mounting the RPC tracking layers onto a metallic support frame designed according to specifications shown in Figure~\ref{fig:proANUBIS_design} using a crane. 
The RPCs underwent performance testing using the ATLAS FELIX DAQ system, confirming their functionality. 
In parallel, the HV and LV systems from CAEN were tested and connected to the DAQ system. 
The DAQ chain was then tested separately and independently, \ie without RPCs, with the data signals generated by function generators. 
Both systems, the RPC detectors and the DAQ setup, were demonstrated to be working well independently from one another. 

\begin{figure}[ht]
\centering
 \begin{subfigure}[b]{\linewidth}
 \centering
\includegraphics[width=6.5cm]{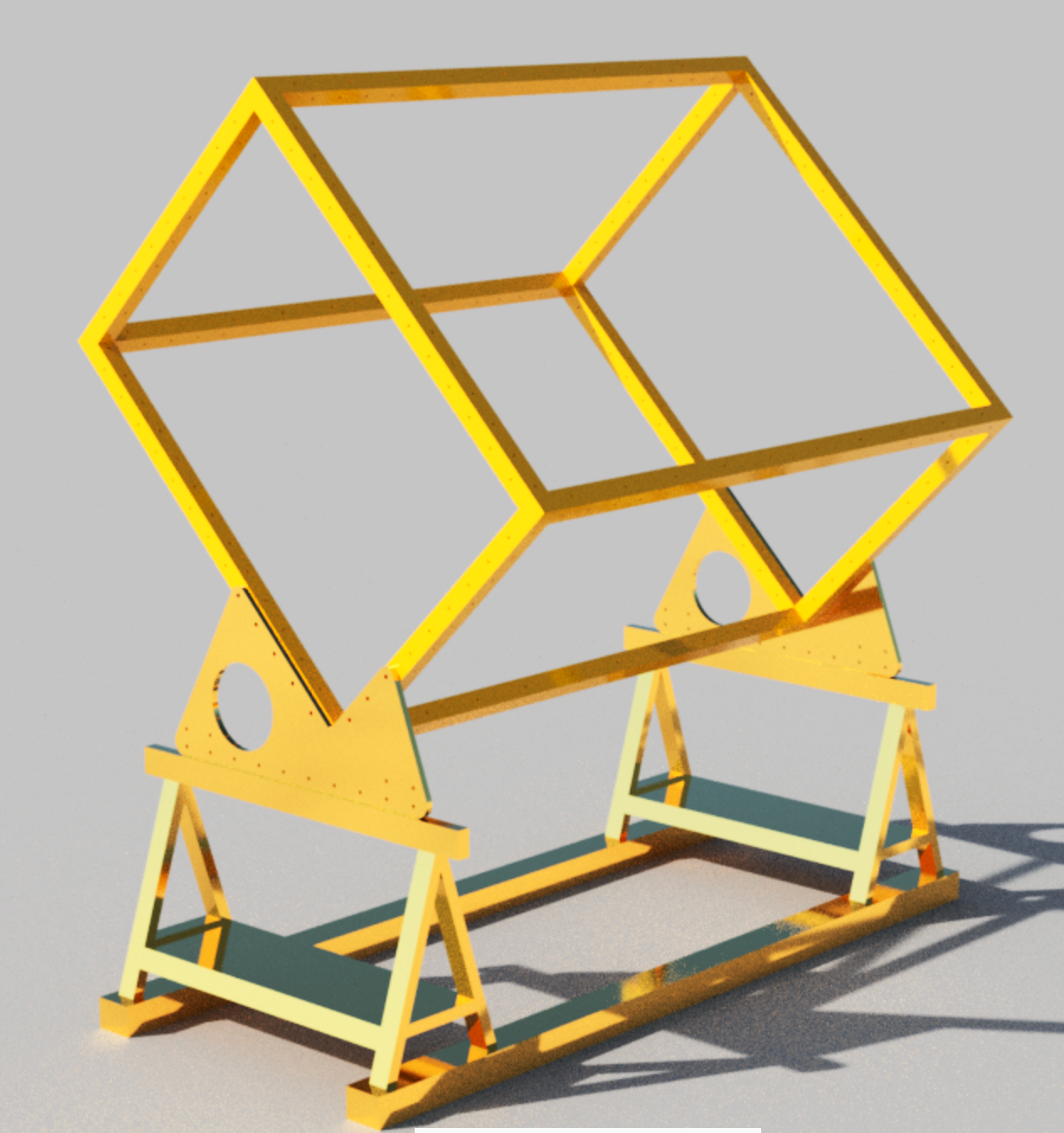}
 \caption{}
 \label{fig:proANUBIS_construction_Top}
\end{subfigure}
 \begin{subfigure}[b]{\linewidth}
  \centering
\includegraphics[width=7 cm]{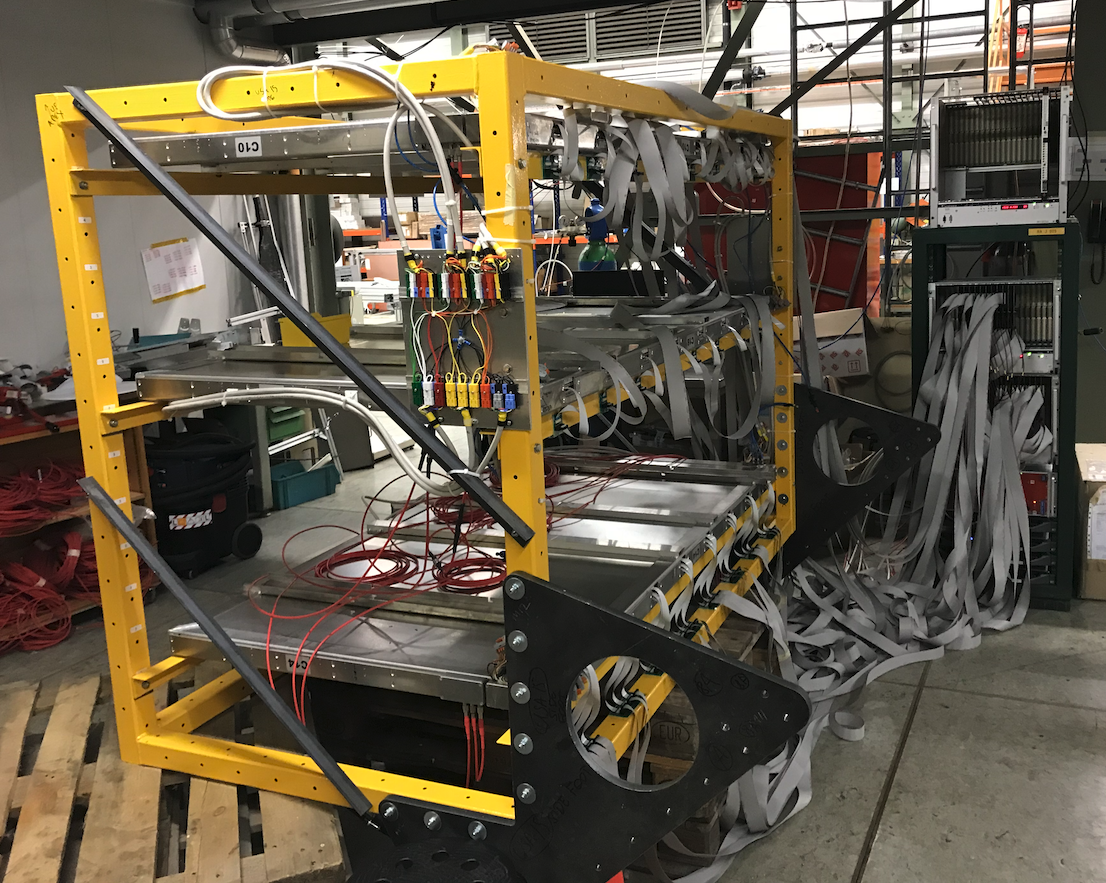}
 \caption{}
 \label{fig:proANUBIS_construction_Low}
\end{subfigure}
\caption{
(a) The design of the steel support frame for the \proanubis detector with detector planes inclined approximately by 45$^{\circ}$ to maximise the geometrical acceptance to particles coming from the ATLAS interaction point. 
(b) The \proanubis steel support frame together with the installed RPC tracking layers in a test of  the DAQ chain at the CERN BB5 laboratory and before deployment in the UX1 underground experimental cavern of the ATLAS experiment.} 
\label{fig:proANUBIS_construction}
\end{figure}

In line with the LHC schedule, the \proanubis setup (Figure~\ref{fig:proANUBIS_construction}), including RPCs and DAQ chain, was transported to Point 1 of the CERN LHC in January 2023, where it was installed in February 2023 (YETS-2022). 
The primary objectives of the \proanubis detector during its operation are to evaluate the detector's performance under representative conditions, synchronisation with ATLAS and reconstruction of muons by both ATLAS and ANUBIS detectors. 
The ultimate objective is the measurement of background rates in background-enriched kinematic regions under realistic conditions present just below the ceiling of the UX1 ATLAS underground cavern, where the future ANUBIS detector is planned to be installed.
The insights gained from this experience proved invaluable for refining the design of the ANUBIS detector.

\vspace{-0.3cm}
\section{Conclusion and outlook} \label{Sec:Summary}
\vspace{-0.2cm}
The ANUBIS experiment aims to explore the parameter space of neutral LLPs at the LHC. 
To evaluate the technical feasibility of this concept, a small-scale detector prototype, \proanubis, was designed and installed in the UX1 underground cavern of the ATLAS experiment. 
The main objectives of \proanubis are providing a proof-of-principle of the detector concept in terms of performance, and the measurement of background rates in background-enriched kinematic regions under experimental conditions that are representative of the full-scale ANUBIS detector.

This report presented the design, construction, and commissioning of \proanubis with particular emphasis on the use of RPCs as the core detector technology. 
The \proanubis setup consists of three tracking layers employing BIS7-type RPCs arranged in triplet, singlet, and doublet configurations from bottom to top. 
A hybrid DAQ system, combining commercial and custom-designed components, was implemented for signal processing and data collection. 
Commissioning tests were conducted at CERN's BB5 laboratory and validated the performance of individual components and the integrated detector system prior to installation. 
Performance characterisation using cosmic ray muons confirmed the expected detection efficiency and operational stability of the RPCs.

With the successful construction and commissioning of \proanubis, the experiment was deployed within the ATLAS cavern and has recorded 104~\fb of $pp$ collision data in 2024 and 73~\fb in 2025. 
These measurements will provide empirical input for refining simulation models, optimising detector layout, and developing robust background rejection strategies for the ANUBIS experiment. 
The experience gained from \proanubis will inform future improvements in mechanical integration, timing performance, and DAQ optimisation, paving the way for the realisation of the project in upcoming LHC runs.

\section*{Acknowledgements}
\vspace{-0.2mm}
We thank our ATLAS colleagues at CERN for their assistance throughout the assembly and testing phases of the individual \proanubis components. 
As well as our CODEX-b colleagues for their temporary loan of a set of FE boards, which was essential to allow for the swift construction and installation of \proanubis. 
The successful commissioning and installation of this project would not have been achievable without the dedicated efforts and collaboration of each individual involved. 

%\clearpage 
\printbibliography[title=References,heading=bibintoc]

\end{document}

%% file: QC_paper/style.tex
\newcommand{\eg}{\mbox{\itshape e.g.}\xspace}
\newcommand{\ie}{\mbox{\itshape i.e.}\xspace}

\newcommand{\mum}{\ensuremath{\mu\textnormal{m}}\xspace}
\newcommand{\mm}{\ensuremath{\textnormal{mm}}\xspace}
\newcommand{\cm}{\ensuremath{\textnormal{cm}}\xspace}
\newcommand{\metre}{\ensuremath{\textnormal{m}}\xspace}

\newcommand{\fb}{\ensuremath{{\rm fb}^{-1}}\xspace}

\def\ns{\text{ns}\xspace}
\def\ps{\text{ps}\xspace}
\def\metre{\text{m}\xspace}

\newcommand{\proanubis}{\textsl{pro}ANUBIS\xspace}

%% file: QC_paper/authorList.tex
% authors (by surname)
%\author[20]{Thomas Adolphus}
\author[18]{Giulio Aielli}
%\author[8,20]{Jahanzeb Akhtar}
%\author[20]{Arun Atwal}
%\author[5]{Martin Bauer?}
%\author[13,20]{Rachel Bentham}
\author[1]{Oleg Brandt}
%\author[20]{Jude Burling}
\author[20]{Jon Burr}
%\author[20]{Patrick Collins}
%\author[3]{Louie Dartmoor Corpe}
%\author[20]{Matthew Coxon}
%\author[1]{Jonas Dej}
%\author[23]{Sofie Nordahl Erner}
%\author[1]{Cayetano Fernandez Ruiz}
%\author[4,7,20]{Jindrich Jelinek}
\author[17]{Oliver Kortner}
\author[17]{Hubert Kroha}
%\author[19,22]{Lawrence Lee Jr}
\author[1]{Christopher Lester}
%\author[1]{Yingshan Liang}
%\author[20]{Kaijia Liu}
%\author[1]{Anna Mullin?}
%\author[16]{Christian Ohm?}
%\author[20]{David Peng}
\author[9]{Luca Pizzimento}
\author[2]{Ludovico Pontecorvo}
%\author[17]{Giorgia Proto}
%\author[12,20]{Th\'eo Reymermier}
\author[1]{Michael Revering}
%\author[21]{Elisa Ritz}
\author[14,15,20]{Thomas P. Satterthwaite}
\author[1]{Aashaq Shah}
%\author[11]{Sinem Simsek}
\author[17]{Daniel Soyk}
%\author[8,20]{Tom Spencer}
\author[1]{Paul Swallow}
\author[]{\newline for the ANUBIS Collaboration}
%\author[20]{Balint Szepfalvi}
%\author[7,20]{Noshin Tarranum}
%\author[10,20]{Olivia Valentino}
%\author[1]{Julian Wack}
%\author[6,20]{Yanglin Wan}
%\author[20]{Peng Wang}
%\author[20]{Esperanza Winter Lopez}
%\author[20]{Monami Yoshioka}
%\author[1]{Yingchang Zhang}

%have to arrange this one in alphabetical order.
\address[1]{Cavendish Laboratory, University of Cambridge, Cambridge, United Kingdom}
\address[2]{CERN, Geneva, Switzerland}
%\address[3]{Universit\'e Clermont Auvergne / Laboratoire de Physique de Clermont Auvergne, CNRS/IN2P3, France}
%\address[4]{Institute of Experimental and Applied Physics, CTU in Prague, Czech Republic}
%\address[5]{IPPP, Dept of Physics, Durham University, Durham, United Kingdom}
%\address[6]{School of Physics and Astronomy, University of Edinburgh, Edinburgh, United Kingdom}
%\address[7]{Department of Nuclear and Particle Physics, University of Geneva, Geneva, Switzerland}
%\address[8]{School of Sciences and Engineering, Glasgow Caledonian University, Glasgow, United Kingdom}
\address[9]{Department of Physics, University of Hong Kong, Hong Kong SAR, China}
%\address[10]{Imperial College London, London, United Kingdom}
%\address[11]{Istinye University, Istanbul, Turkey}
%\address[12]{Institut de Physique des 2 Infinis (IP2I), Lyon, France}
%\address[13]{Department of Physics and Astronomy, University of Manchester, Manchester, United Kingdom}
\address[14]{Department of Physics, Stanford University, Stanford, CA 94305, USA}
\address[15]{Kavli Institute for Particle Astrophysics and Cosmology, Stanford, CA 94305, USA}
%\address[16]{KTH Royal Institute of Technology, Stockholm, Sweden}
\address[17]{Max Planck Institute for Physics, Munich, Germany}
\address[18]{University and INFN Roma Tor Vergata, Rome, Italy}
%\address[19]{Department of Physics, University of Tennessee, Knoxville, TN 37996, USA}
\address[20]{Formerly at: Cavendish Laboratory, University of Cambridge, Cambridge, United Kingdom}
%\address[21]{Formerly at: Hamburg University, Hamburg, Germany}
%\address[22]{Formerly at: Department of Physics, Harvard University, Cambridge, USA}
%\address[23]{Formerly at: IPPP, Dept of Physics, Durham University, Durham, United Kingdom}